\documentclass[%
 reprint,
 amsmath,amssymb,
 aps,
pra,
]{revtex4-2}

\usepackage{graphicx}
\usepackage{dcolumn}
\usepackage[table]{xcolor}
\usepackage[normalem]{ulem}

\usepackage{bm}

\begin{document}

\preprint{APS/123-QED}

\preprint{APS/123-QED}

\title{Features of the van der Waals Interaction on the Cesium $6S_{1/2} \rightarrow 7P_{3/2}$ Transition in an Optical Nanocell}

\author{Armen Sargsyan}\thanks{}
\altaffiliation{}
\affiliation{Institute for Physical Research, National Academy of Sciences of Armenia, 0204 Ashtarak, Armenia}
\mbox{}
\author{Anahit Gogyan}
\email{agogyan@gmail.com}
\affiliation{Institute for Physical Research, National Academy of Sciences of Armenia, 0204 Ashtarak, Armenia}
\mbox{}
\author{David Sarkisyan}
\affiliation{Institute for Physical Research, National Academy of Sciences of Armenia, 0204 Ashtarak, Armenia}
\mbox{}

\date{\today}

\begin{abstract}
We report the first experimental study of the influence of a dielectric surface on the transmission spectrum of cesium atoms for the $6S_{1/2} \rightarrow 7P_{3/2}$ ($D_2$) transition in vapor cells with thicknesses in the range $50$--$250\,$nm. The measurements were performed using a homemade optical nanocell filled with atomic cesium and featuring a wedge-shaped gap between the inner surfaces of sapphire windows. For atom--surface distances below approximately $300$\,nm, a significant red shift of the atomic transition frequency is observed due to van der Waals (vdW) interactions with the dielectric surface. An additional red shift arises at high vapor pressures owing to Cs--Cs interactions, with a measured contribution of $15$\,MHz/Torr, which must be accounted for in order to correctly determine the vdW coefficient $C_3$. By recording transmission spectra of the nanocell at different thicknesses, we determine for the first time the vdW coefficient for the Cs $6S_{1/2} \rightarrow 7P_{3/2}$ transition, obtaining values in the range $C_3 \sim 2- 20$\,kHz\,$\mu$m$^3$. These results are of interest for fundamental studies of atom--surface interactions and are also relevant for the development of miniature sensors based on atomic vapors, in particular compact frequency references exploiting atomic transitions in the blue spectral region.
\end{abstract}
\maketitle

\section{Introduction}
\label{intro}

Cesium and rubidium atomic vapors continue to be widely employed in both fundamental research and practical applications involving optical and magneto--optical processes~\cite{1,2,3,4,5}. This widespread use is primarily caused by the presence of strong atomic transitions in the near--infrared spectral range ($700$--$900\,\mathrm{nm}$), where stable and well--developed laser sources are readily available. In addition, the preparation of these elements in the form of a vapor column confined within spectroscopic cells of centimeter-scale thickness is technically straightforward. Even at room temperature, the atomic vapor density reaches $\sim 10^{10}\,\mathrm{cm^{-3}}$, which facilitates reliable recording of absorption and fluorescence spectra. A comprehensive review of sensor devices employing micron-thick vapor cells is presented in Ref.~\cite{6}.

For practical applications, further miniaturization of spectroscopic cells is highly desirable. However, as the cell dimensions are reduced, the average distance between atoms and the cell windows correspondingly decreases, and atom--surface interactions begin to play a significant role. Early investigations of these effects were based on analysis of the line shape of selective reflection (SR) of laser radiation from the inner surface of a glass cell window in contact with atomic vapor~\cite{7}. The SR signal was originally attributed to atoms located within a characteristic distance $L_{\text{SR}}\sim \lambda/2\pi$ from the surface, where $\lambda$ is the wavelength of the resonant atomic transition. Subsequent studies~\cite{8,9}, however, indicated that the SR signal may be formed by atoms traversing distances exceeding $\lambda/2\pi$, implying that the corresponding atomic time of flight is comparable to the spontaneous decay time.

The atom--surface interaction in the non--retarded regime is commonly referred to as the van der Waals (vdW) interaction, also known as the Casimir--Polder interaction~\cite{10,11,12}. A qualitative picture of this interaction is provided in Ref.~\cite{13}, where an atom is modelled as an electric dipole interacting with its mirror image induced in the dielectric medium of the cell window. Although the time--averaged electric dipole moment of an atom is zero, its instantaneous value fluctuates~\cite{7}, giving rise to an electric field due to the induced image dipole. This field modifies the atomic energy levels, resulting in frequency shifts and spectral broadening of atomic transitions. The magnitude of this effect increases with the principal quantum number $n$, as higher--lying electronic states are more weakly bound to the nucleus and therefore more strongly influenced by the induced dipole in the dielectric. Consequently, one expects the vdW frequency shift for the $6S_{1/2}\!\rightarrow\!7P_{3/2}$ transition to exceed that of the $6S_{1/2}\!\rightarrow\!6P_{3/2}$ transition.

Recently, optical nanocells (NCs) filled with alkali--metal vapors have proven to be particularly well suited for investigating atom--surface vdW interactions, as they allow atom--surface separations on the order of tens of nanometers. Using techniques such as resonant absorption~\cite{12,13,14}, selective reflection from nanocells~\cite{13,15,16}, and resonant fluorescence~\cite{17}, vdW interaction coefficients $C_3$ have been measured for Cs atoms at $\lambda = 852$ and $895\,\mathrm{nm}$ (D$_1$ and D$_2$ lines) and for Rb atoms at $\lambda = 780$ and $795\,\mathrm{nm}$ (D$_1$ and D$_2$ lines). For Cs atoms at 895 and $852\,\mathrm{nm}$, the reported vdW coefficients are $C_3 = (1 \pm 0.2)\,\mathrm{kHz}\,\mu\mathrm{m}^3$ and $C_3 = (2 \pm 0.2)\,\mathrm{kHz}\,\mu\mathrm{m}^3$ for the D$_1$ and D$_2$ lines, respectively~\cite{7}.

In the present work, we investigate for the first time the characteristics of the vdW interaction for the Cs $6S_{1/2}\!\rightarrow\!7P_{3/2}$ transition by recording transmission spectra of $456\,\mathrm{nm}$ radiation through a nanocell with cesium vapor column thicknesses in the range $50$--$250\,\mathrm{nm}$. A comparative analysis of the vdW--induced spectral features of the $6S_{1/2}\!\rightarrow\!6P_{3/2}$ and $6S_{1/2}\!\rightarrow\!7P_{3/2}$ transitions is also presented. The spectral density, corresponding to the peak of the vdW--modified transmission spectrum, is maximal for atoms located near the center of the nanocell, $z=L/2$, where $z$ denotes the distance from a nanocell window. For these atoms, the vdW frequency shift is minimal, while the spectral contribution is maximal~\cite{13}. The red frequency shift arising from the simultaneous influence of both nanocell windows is given by
\begin{equation}
\Delta \nu_{\mathrm{vdW}} = -\frac{16 C_3}{L^3},
\label{eq:vdW_shift}
\end{equation}
where $C_3$ is the vdW interaction coefficient, defined as the difference between the $C_3$ coefficients of the excited and ground states~\cite{7,13,18}.

\begin{figure}[htb]
    \centering
    \includegraphics[width=0.5\textwidth]{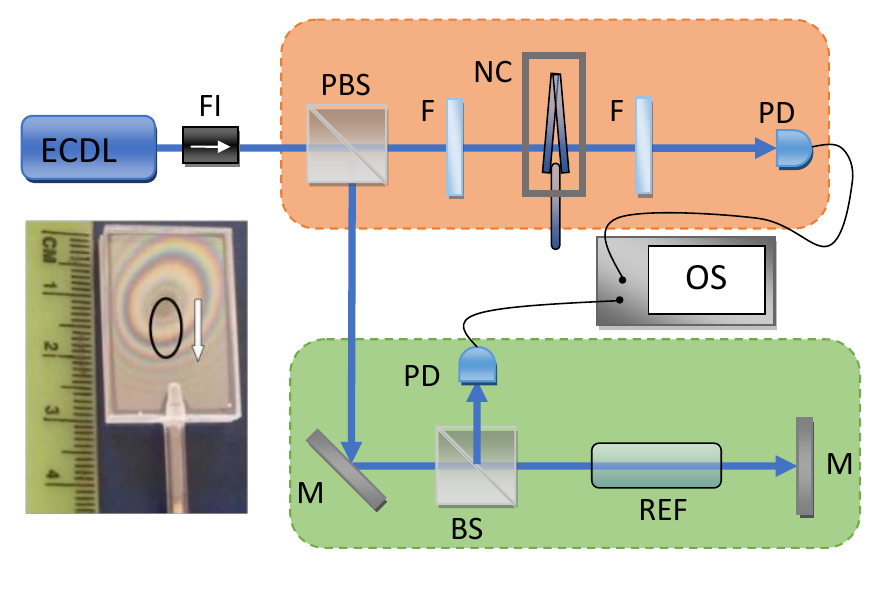}
    \caption{Scheme of the experimental setup. 
ECDL~-- tunable extended-cavity diode lasers operating at 852 and 456~nm; 
FI~-- Faraday isolator; 
PBS~-- polarizing beam splitter; 
BS~-- beam splitter; 
F~-- optical glass filters; 
NC~-- optical nanocell (NC) filled with cesium vapor, with windows made of technical sapphire; 
PD~-- photodetectors; 
REF~-- reference cell for saturated absorption (SA) spectroscopy; 
OS~-- digital oscilloscope. 
The orange area inset illustrates the measurement the nanocell gap thickness. 
On the left, a photograph of the NC is shown; the oval marks the region where the gap thickness is $50$--$250\,\mathrm{nm}$, and the vertical white arrow indicates the direction of thickness variation.}
    \label{fig:1}
\end{figure}

\begin{figure}[ht]
    \centering
    \includegraphics[width=0.3\textwidth]{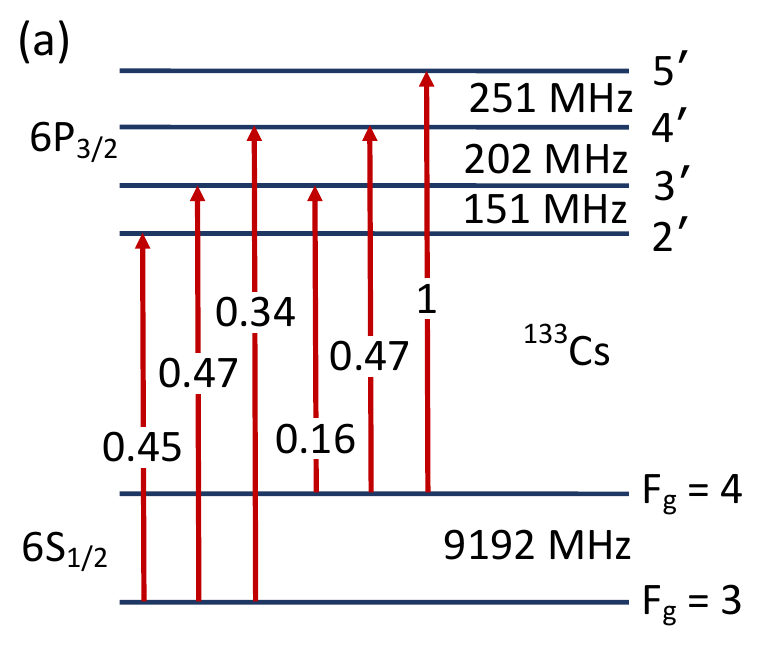}
    \includegraphics[width=0.3\textwidth]{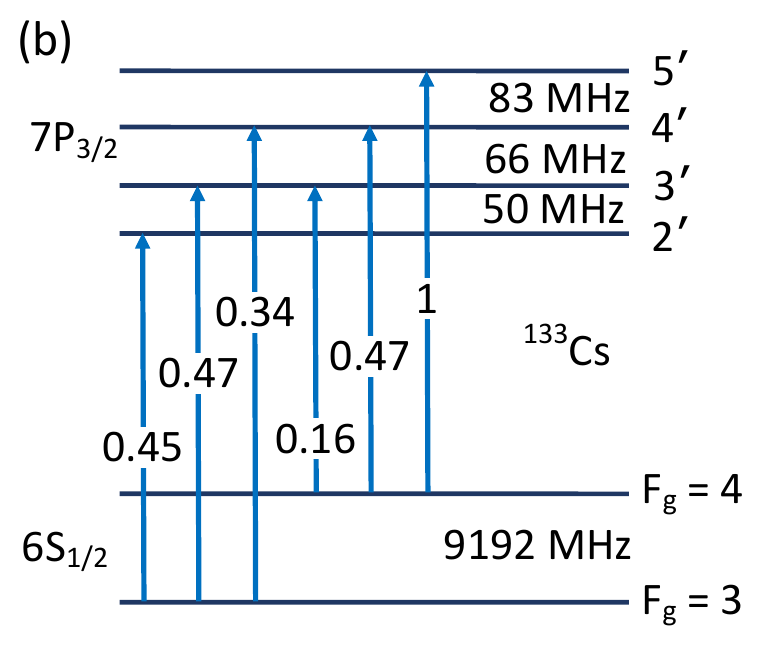}
    \caption{Cesium atomic energy-level diagrams and relative transition strengths. 
The intensity of the strongest transition $4\!\rightarrow\!5'$ is normalized to unity. 
(a) Hyperfine transitions $F_g=3,4 \rightarrow F_e=2',3',4',5'$ of the $6S_{1/2}\!\rightarrow\!6P_{3/2}$ ($D_2$) line at $\lambda=852\,\mathrm{nm}$; upper hyperfine levels are indicated by dashed lines. 
(b) Hyperfine transitions $F_g=3,4 \rightarrow F_e=2',3',4',5'$ of the $6S_{1/2}\!\rightarrow\!7P_{3/2}$ transition at $\lambda=456\,\mathrm{nm}$. 
The oscillator strength of the $6S_{1/2}\!\rightarrow\!7P_{3/2}$ transition is more than 60 times smaller than that of the $6S_{1/2}\!\rightarrow\!6P_{3/2}$ transition. 
Transition frequencies and hyperfine splittings are not drawn to scale.}
    \label{fig:2}
\end{figure}

\begin{figure}[ht]
    \centering
    \includegraphics[width=0.45\textwidth]{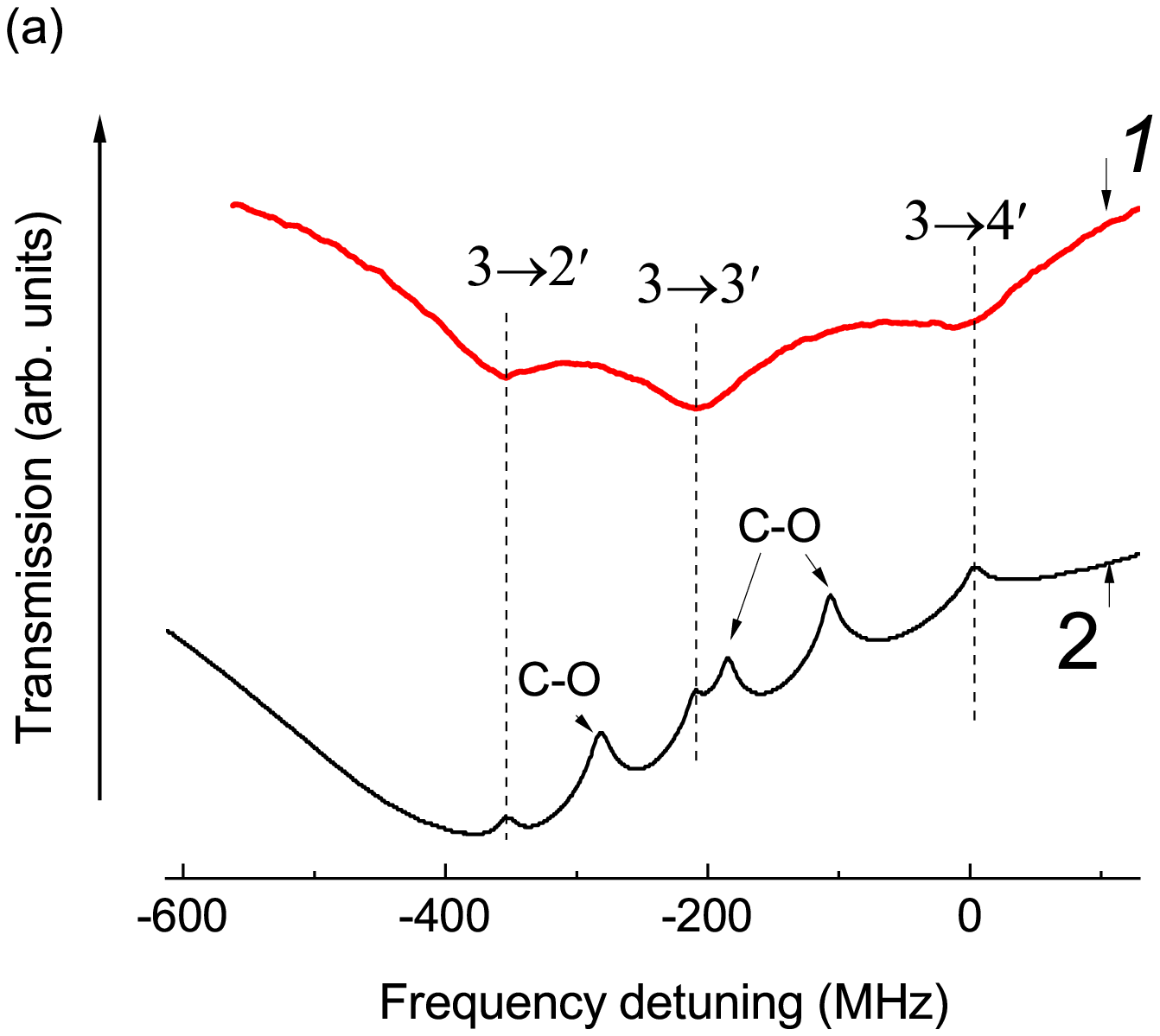}
    \includegraphics[width=0.45\textwidth]{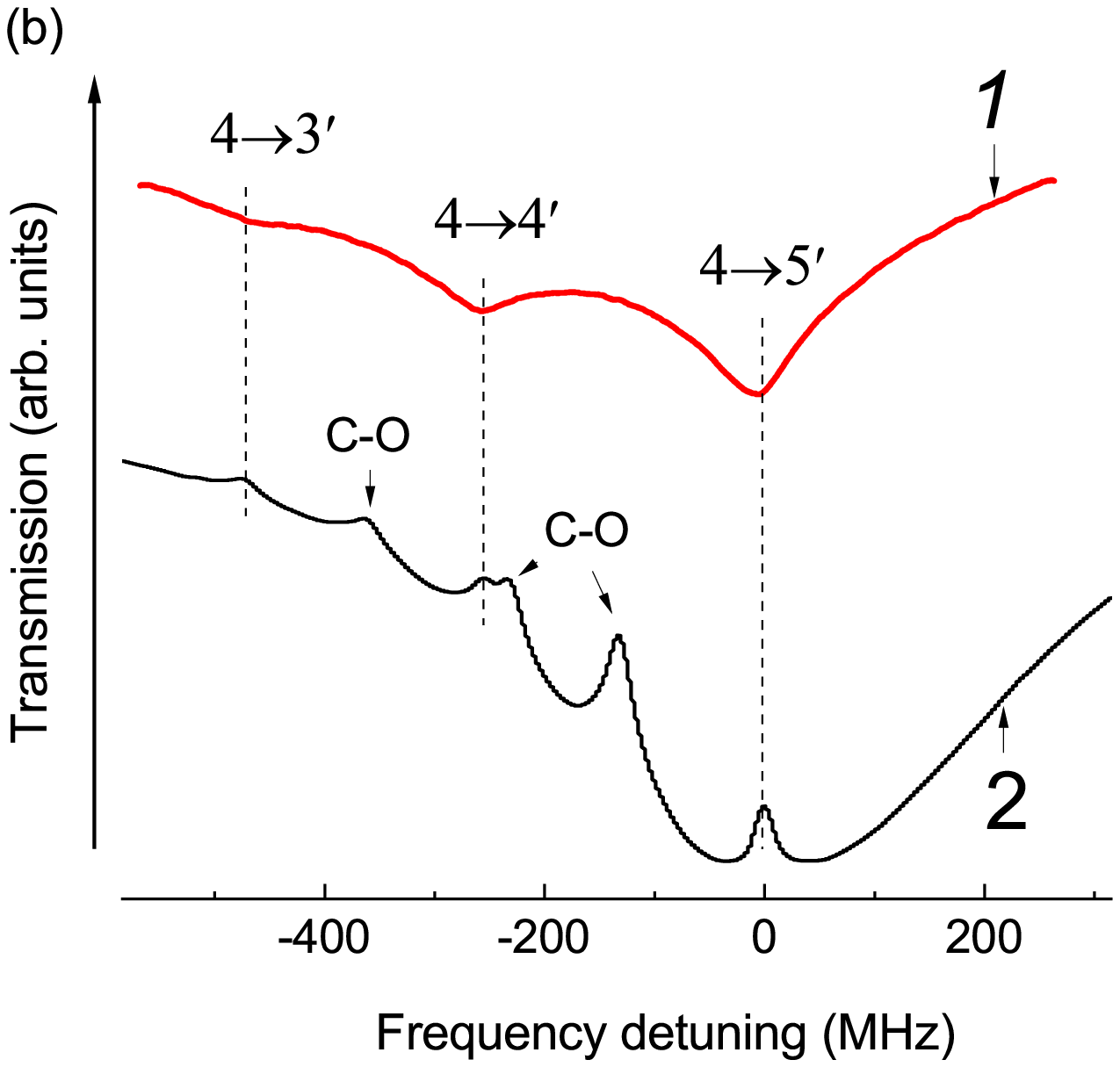}
    \caption{Transmission spectra of cesium vapor in a nanocell for the $6S_{1/2}\!\rightarrow\!6P_{3/2}$ transition at $\lambda=852\,\mathrm{nm}$. 
Upper curves (1) show nanocell transmission spectra for a thickness $L=230\,\mathrm{nm}$; lower curves (2) show the corresponding saturated absorption (SA) reference spectra. 
(a) Transitions $3\!\rightarrow\!2',3',4'$; 
(b) transitions $4\!\rightarrow\!3',4',5'$. 
No measurable frequency shift is observed due to the small value of the van der Waals coefficient $C_3$ for this transition. 
C--O denote the well-known crossover resonances in saturated absorption spectra~\cite{20,21,22}.}
    \label{fig:3}
\end{figure}

\section{Experimental Setup and Results}
\label{sec:discussion}

Figure~1 shows a schematic of the experimental setup. To study the van der Waals (vdW) spectra of the Cs $6S_{1/2}\!\rightarrow\!6P_{3/2}$ transition, a continuous-wave external-cavity diode laser (ECDL) tunable around $\lambda=852\,\mathrm{nm}$ with a linewidth of $0.1\,\mathrm{MHz}$ was used. For investigation of the $6S_{1/2}\!\rightarrow\!7P_{3/2}$ transition, a continuous-wave ECDL tunable around $\lambda=456\,\mathrm{nm}$ with a linewidth of $0.4\,\mathrm{MHz}$ was employed. 
The NC was placed inside a two-section oven that allowed independent temperature control of the windows and the side-arm reservoir containing metallic Cs. The NC has a wedge-shaped geometry providing continuously varying thickness $L$ between the inner surfaces of the NC windows, which defines the length of the atomic vapor column. Detailed descriptions of the NC design are given in Refs.~\cite{15,16,17}.  The oval in Fig.~1 highlights the region of particular interest, corresponding to $L=50$--$250\,\mathrm{nm}$. Vertical translation of the NC enabled laser radiation to pass through regions of different gap thickness. The thickness $L$ was determined using the method described in Ref.~\cite{19}.  The atomic vapor density was determined by the reservoir temperature $T$.

\begin{figure}[ht]
    \centering
    \includegraphics[width=0.4\textwidth]{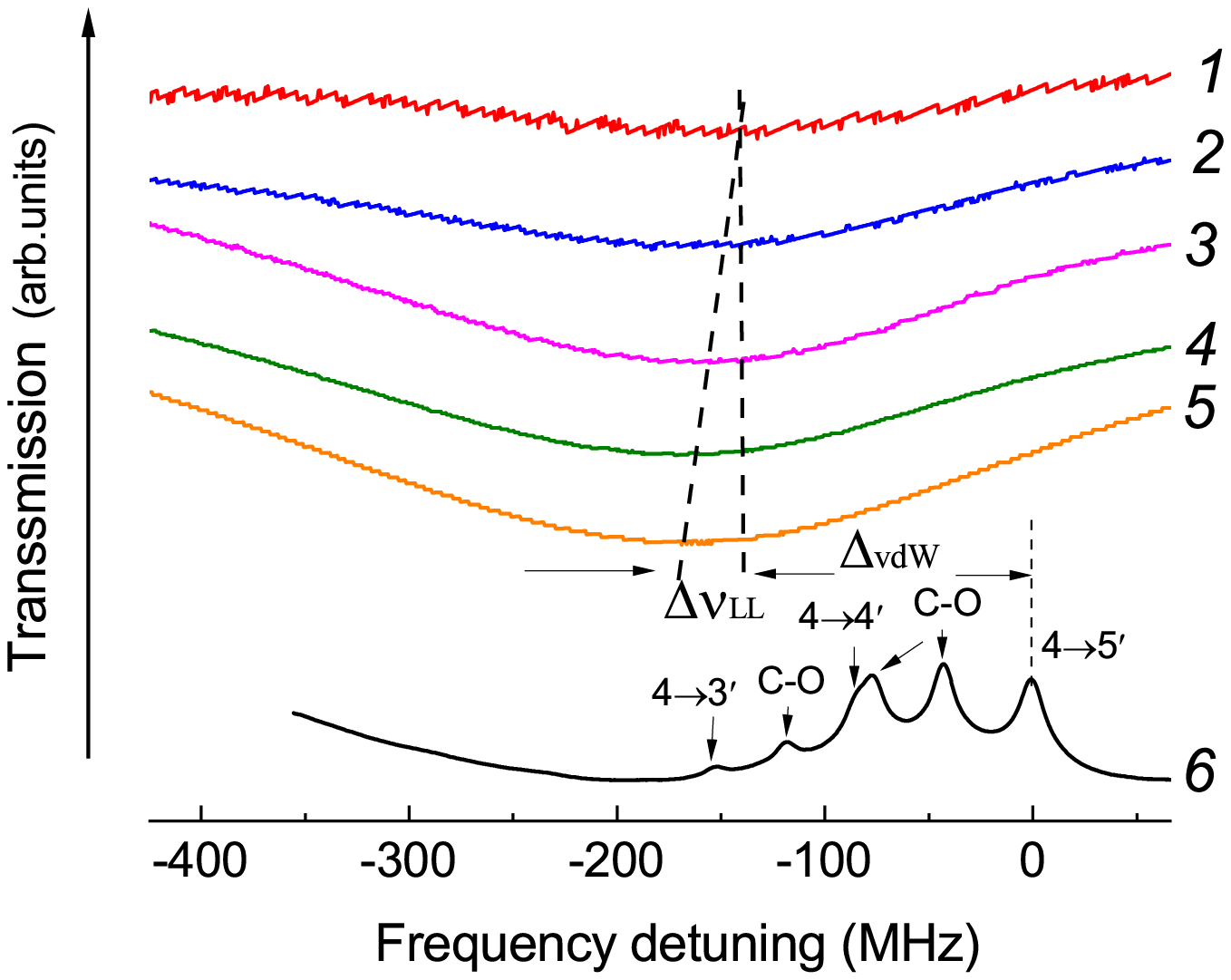}
    \caption{Transmission spectra of cesium vapor at $\lambda=456\,\mathrm{nm}$ for the $6S_{1/2}\!\rightarrow\!7P_{3/2}$ transitions $4\!\rightarrow\!3',4',5'$ as a function of vapor temperature (pressure). 
Curves 1--5 correspond to temperatures of 200, 245, 270, 295, and $310\,^{\circ}\mathrm{C}$, respectively. 
The nanocell thickness is $L=100\pm5\,\mathrm{nm}$ and the laser power is $3\,\mathrm{mW}$. 
The pressure-induced frequency shift is $\Delta\nu_{\mathrm{LL}}\approx15\,\mathrm{MHz/Torr}$. 
The lower curve (6) shows the SA reference spectrum of the corresponding transitions.}
    \label{fig:4}
\end{figure}

\begin{figure}[ht]
    \centering
    \includegraphics[width=0.4\textwidth]{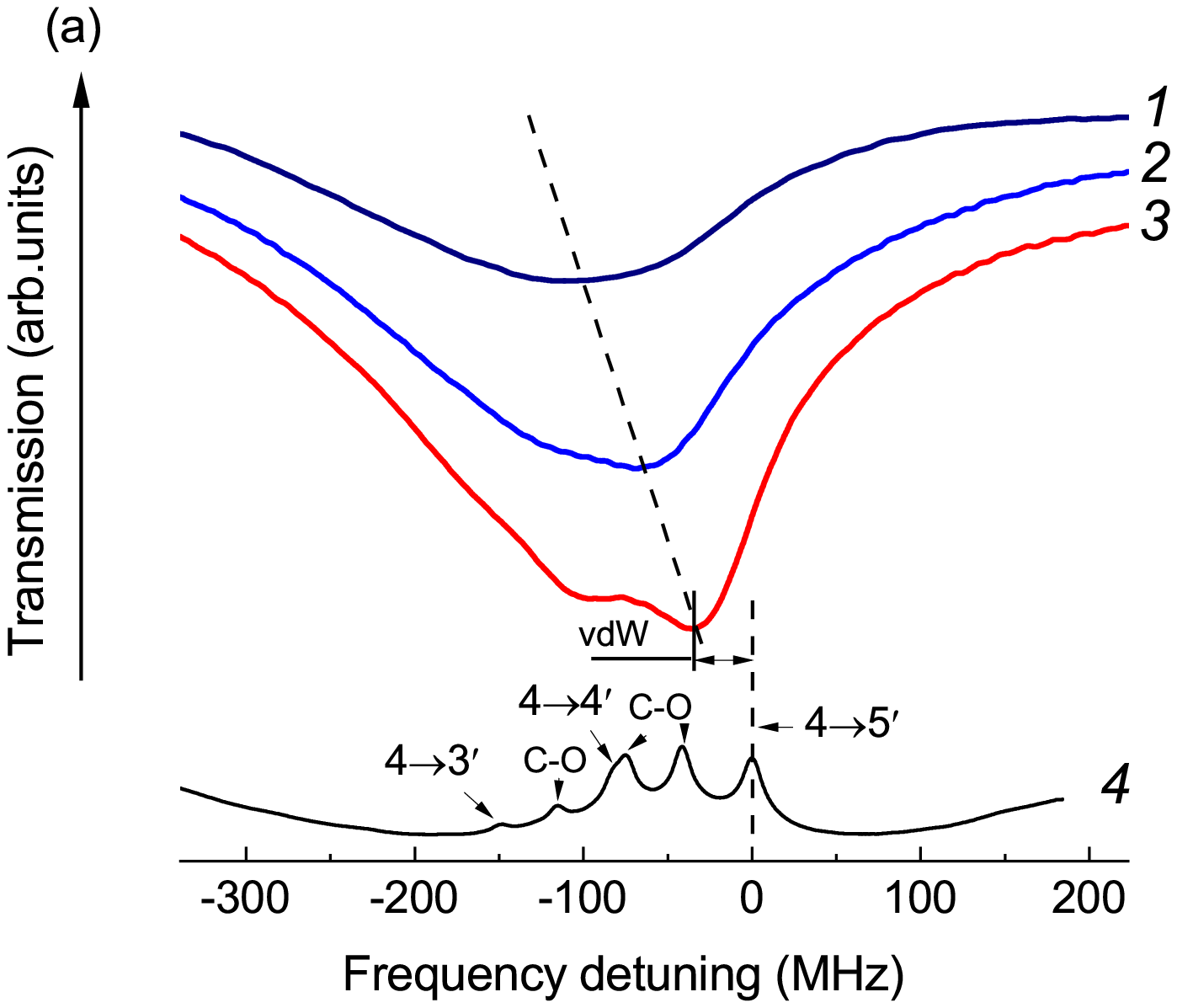}
    \includegraphics[width=0.4\textwidth]{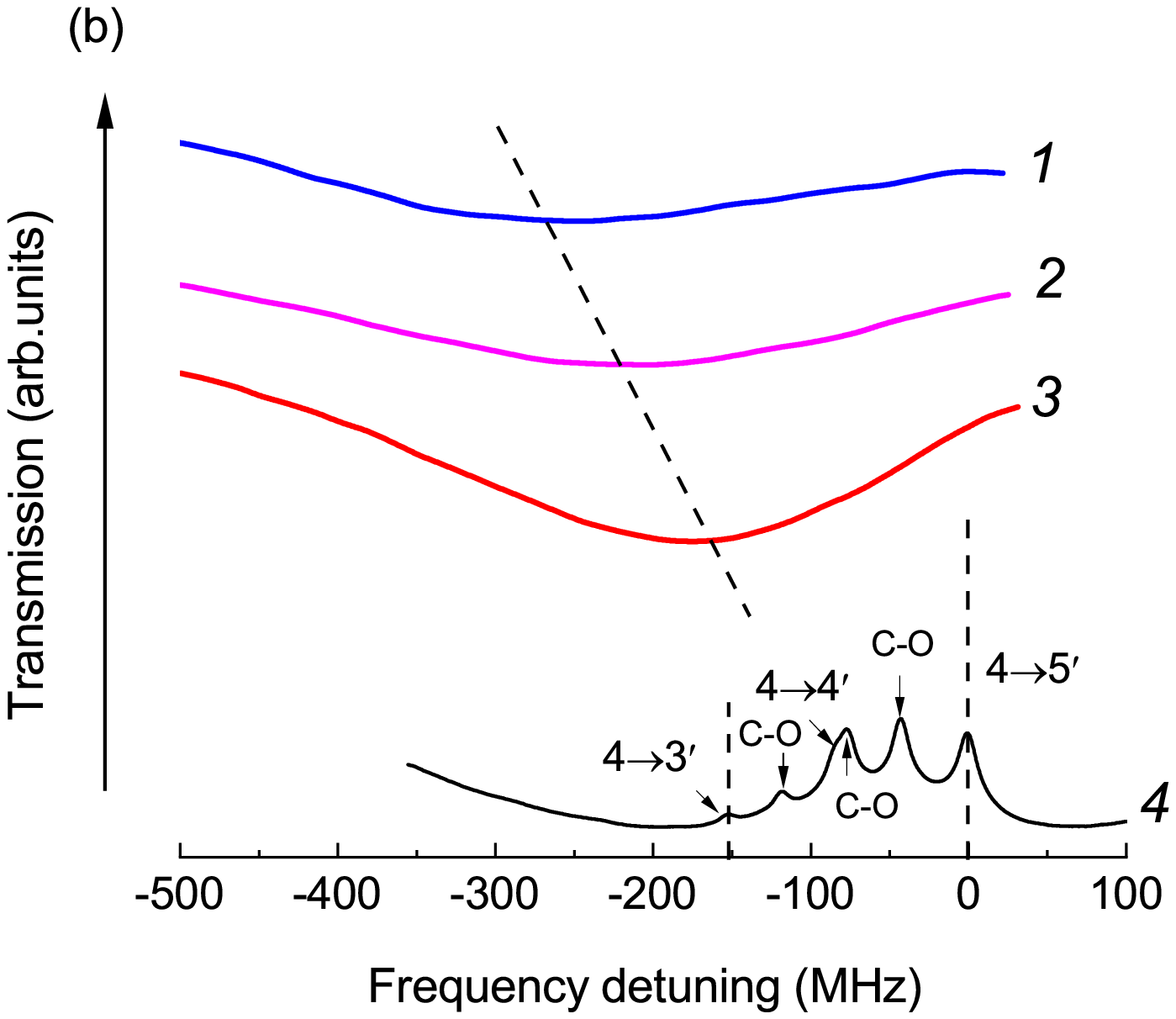}
    \caption{Transmission spectra of the Cs $6S_{1/2}\!\rightarrow\!7P_{3/2}$ transition for different nanocell thicknesses. 
(a) Curves 1--3 correspond to thicknesses $L=140\pm5$, $190\pm5$, and $230\pm5\,\mathrm{nm}$, respectively, at a nanocell temperature of $190\,^{\circ}\mathrm{C}$. 
(b) Curves 1--3 correspond to thicknesses $L=55\pm5$, $65\pm5$, and $100\pm5\,\mathrm{nm}$, respectively, at a temperature of $230\,^{\circ}\mathrm{C}$. 
Bottom curves (4) show the SA reference spectra for the corresponding transitions.}
\label{fig:5}
\end{figure}

\begin{figure}[ht]
    \centering
    \includegraphics[width=0.5\textwidth]{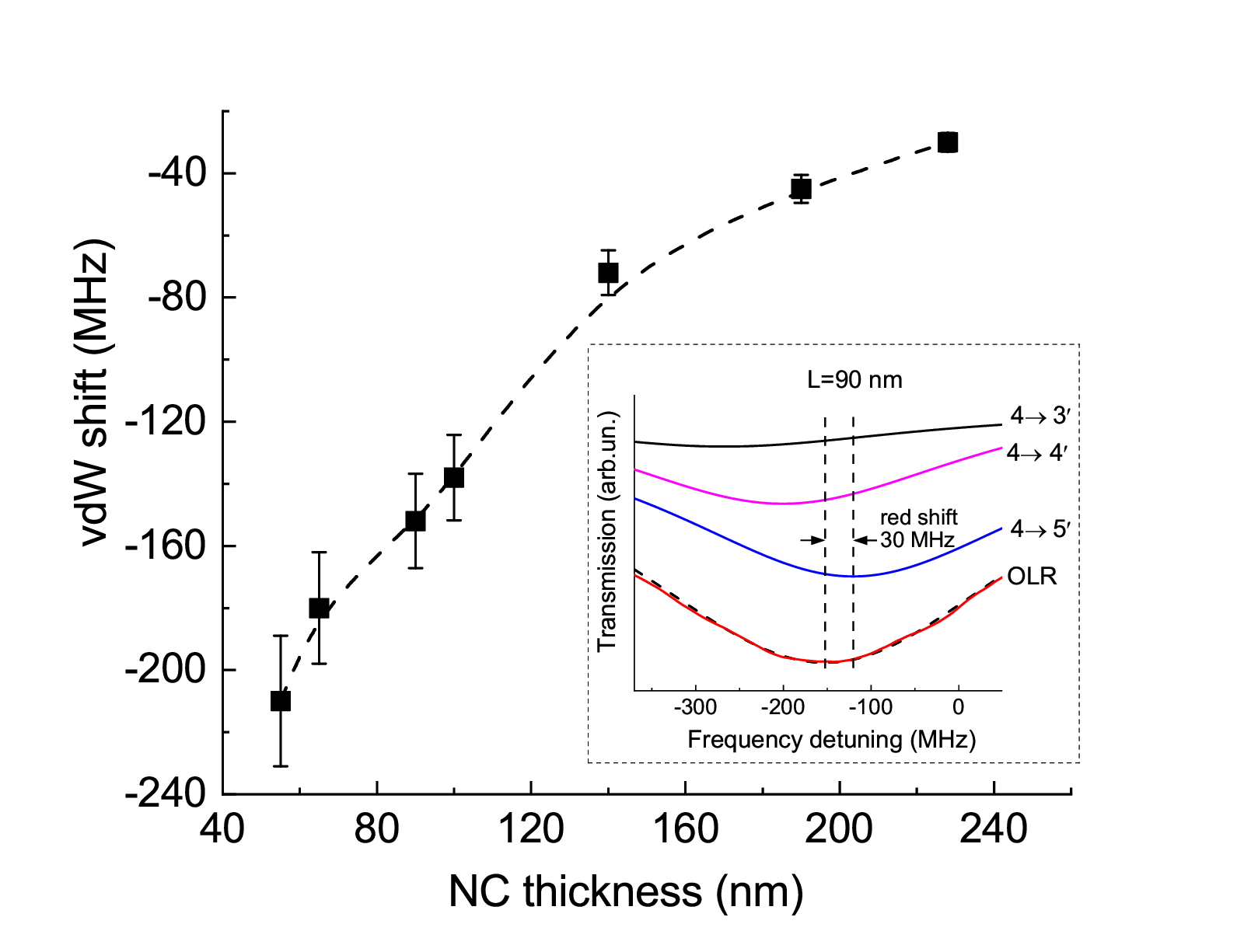}
\caption{Dependence of the van der Waals (vdW) red frequency shift for the $4\!\rightarrow\!5'$ transition of the $6S_{1/2}\!\rightarrow\!7P_{3/2}$ line on the nanocell thickness $L$. 
The inset shows the overlapping red (OLR) shift of approximately $30\,\mathrm{MHz}$, which is subtracted from the total measured shift in order to correctly determine the vdW coefficient $C_3$ using Eq.~(1).}
    \label{fig:6}
\end{figure}

\begin{figure}[ht]
    \centering
    \includegraphics[width=0.5\textwidth]{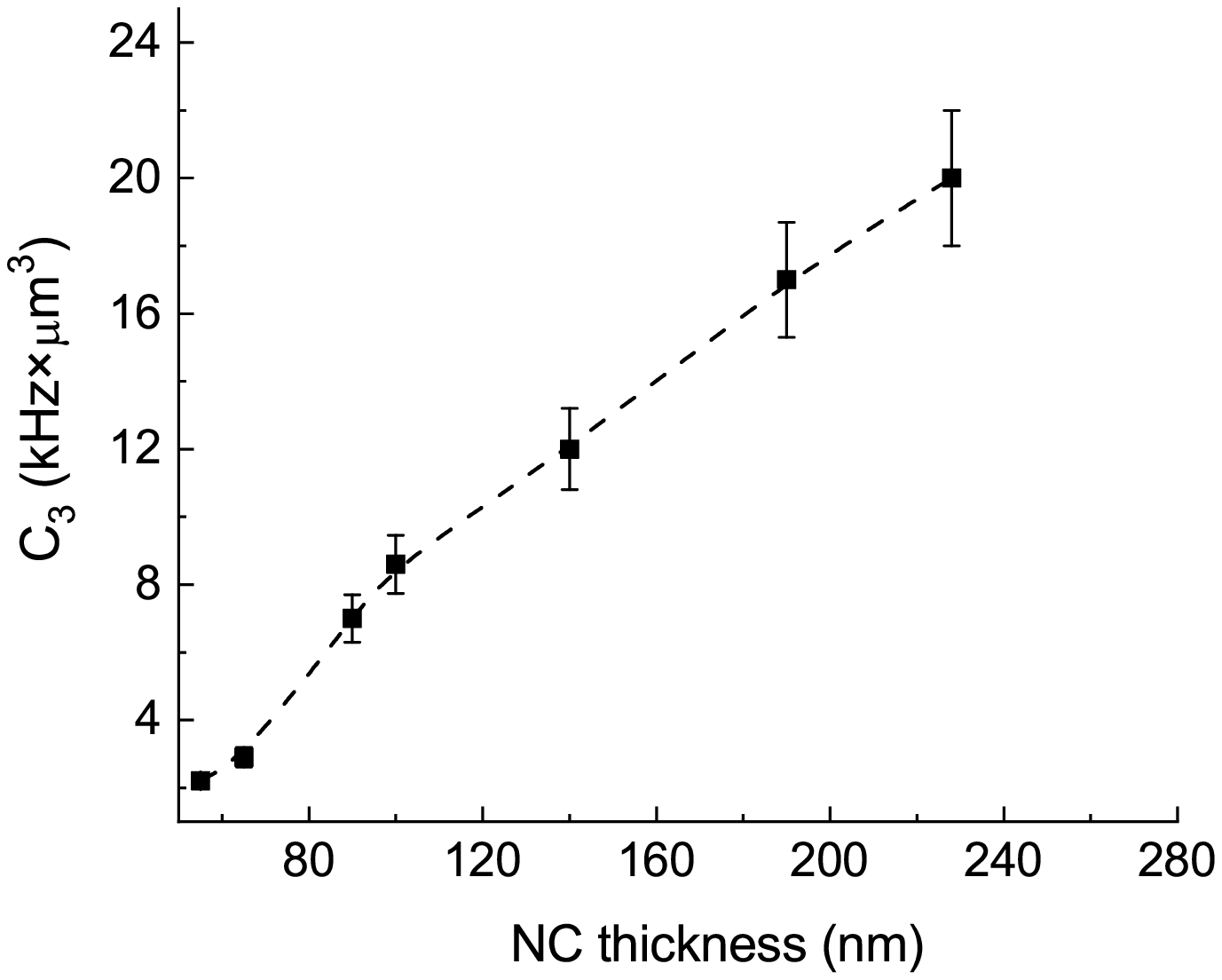}
\caption{Extracted values of the van der Waals coefficient $C_3$ for the $4\!\rightarrow\!5'$ transition as a function of nanocell thickness $L$. 
The overall uncertainty in $C_3$ is estimated to be about 10\%, arising from pressure-induced shifts and from the uncertainty in determining the atom--surface distance $L$ appearing in Eq.~(1).}
    \label{fig:7}
\end{figure}

\begin{figure}[ht]
    \centering
    \includegraphics[width=0.5\textwidth]{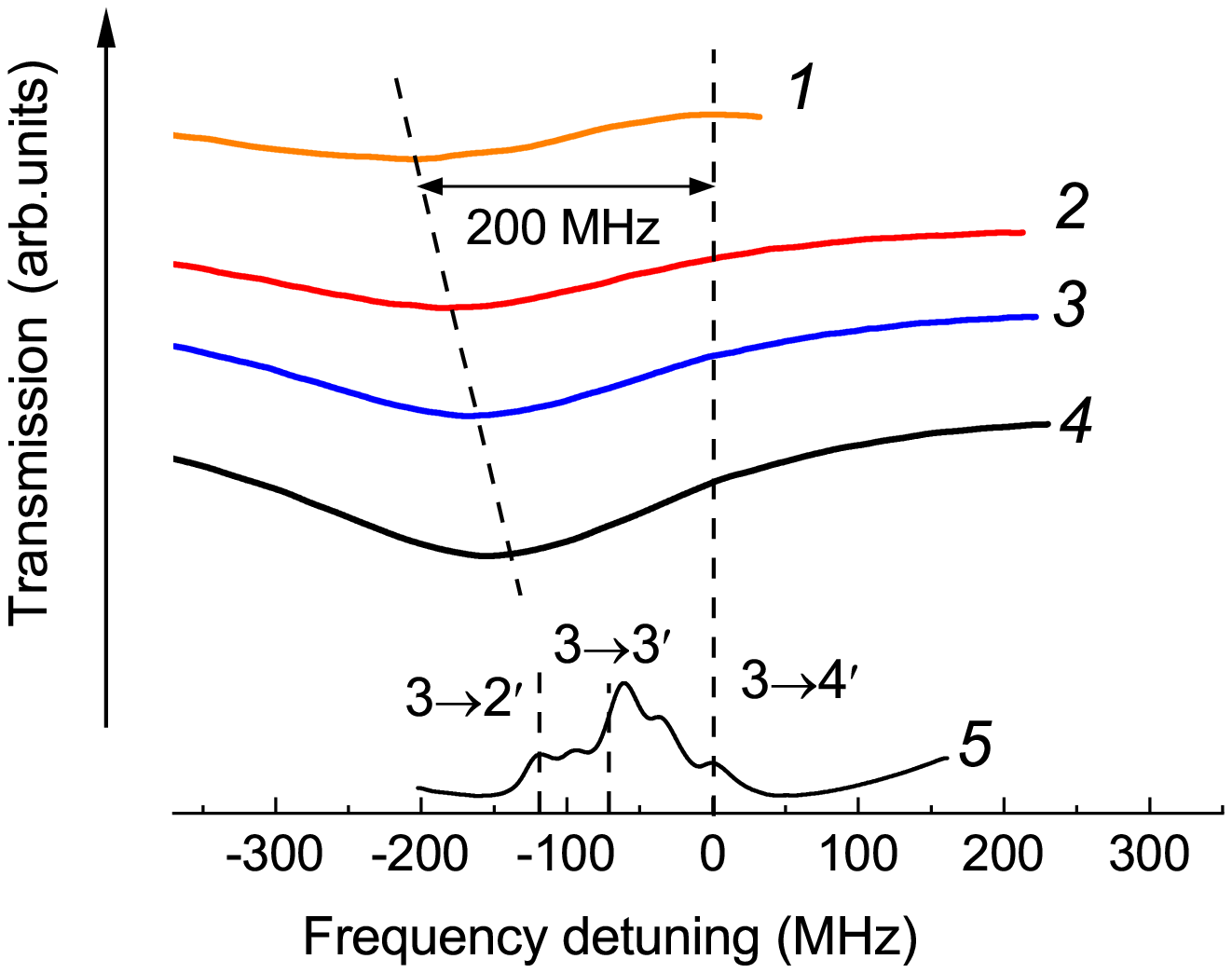}
\caption{Transmission spectra of the Cs $6S_{1/2}\!\rightarrow\!7P_{3/2}$ transition for the hyperfine components $3\!\rightarrow\!2',3',4'$ at different nanocell thicknesses. 
Curves 1--4 correspond to thicknesses $L=60\pm5$, $70\pm5$, $90\pm5$, and $100\pm5\,\mathrm{nm}$, respectively. 
The nanocell temperature is $230\,^{\circ}\mathrm{C}$ and the laser power is $3\,\mathrm{mW}$. 
The bottom curve (5) shows the SA reference spectrum of the corresponding transitions.}
    \label{fig:8}
\end{figure}

Figure~2(a) presents the energy-level diagram of the Cs atom for the $6S_{1/2}\!\rightarrow\!6P_{3/2}$ ($D_2$) transition at $\lambda=852\,\mathrm{nm}$, including the hyperfine transitions $F_g=3,4\rightarrow F_e=2',3',4',5'$ and the hyperfine splittings of the excited $6P_{3/2}$ state. Figure~2(b) shows the corresponding diagram for the $6S_{1/2}\!\rightarrow\!7P_{3/2}$ transition at $\lambda=456\,\mathrm{nm}$. As evident from the diagrams, the frequency spacing between the hyperfine components of the $7P_{3/2}$ state is approximately three times smaller than that of the $6P_{3/2}$ state. The relative transition intensities are also indicated, with the strongest transition $4\rightarrow5'$ normalized to unity. In addition, the vdW interaction coefficient for the $6S_{1/2}\!\rightarrow\!7P_{3/2}$ transition is estimated to be $C_3\sim20\,\mathrm{kHz}\,\mu\mathrm{m}^3$~\cite{14}, which is significantly larger than that for the $6S_{1/2} \rightarrow 6P_{3/2}$ transition. This motivates a direct experimental comparison of vdW-induced frequency shifts for these two transitions starting from 50\,nm.

The laser beam of diameter of 0.6~mm was directed perpendicular to the NC windows, and transmission spectra were recorded while scanning the laser frequency across the atomic resonances. The transmitted radiation was detected using FD--24K photodiodes; the signals were amplified and recorded with a Tektronix TDS2014B four-channel oscilloscope. A frequency reference was obtained by directing a fraction of the laser radiation to a separate unit employing the saturated absorption (SA) technique~\cite{20,21,22} in a centimeter-long Cs cell.

\subsection{Study of the  $6S_{1/2} \rightarrow 6P_{3/2}$ transition}

Figure~3(a), curve~1, shows the transmission spectrum of transitions $6S_{1/2}, F = 3 \rightarrow 6P_{3/2}, F' = 2', 3', 4'$ at $\lambda=852\,\mathrm{nm}$ through a NC with thickness $L\simeq230\pm5\,\mathrm{nm}$ at a cell temperature of $110^{\circ}\mathrm{C}$ and laser power of $7\,\mu\mathrm{W}$. As demonstrated in Refs.~\cite{23,24}, the minimum spectral width of approximately $80\,\mathrm{MHz}$ (FWHM) occurs at $L=\lambda/2$, corresponding to Dicke narrowing. For the $6S_{1/2}\!\rightarrow\!6P_{3/2}$ transition this condition is $L=426\,\mathrm{nm}$, yielding a linewidth roughly five times smaller than the Doppler width at $100^{\circ}\mathrm{C}$. With further reduction of $L$, the linewidth increases due to atom--wall collisions, and for $L\lesssim130\,\mathrm{nm}$ the vdW interaction also contributes to spectral broadening and frequency shifts. Figure~3(b), curve~1, presents the transmission spectrum at $\lambda = 852\,$nm recorded for a nanocell thickness of $L \simeq 230 \pm 5\,$nm, corresponding to the $4 \rightarrow 3',4',5'$ hyperfine transitions.
Owing to the relatively small $C_3$ coefficient for this transition, the frequency shift at $L=230\,\mathrm{nm}$ is negligible compared to the reference cell. The vdW-induced broadening and shifts for $L<100\,\mathrm{nm}$ are discussed in detail in Ref.~\cite{25}.

\subsection{Study of the $6S_{1/2} \rightarrow 7P_{3/2}$ transition}

The study of the $6S_{1/2}\!\rightarrow\!7P_{3/2}$ transition presents several experimental challenges. First, the hyperfine splitting of the $7P_{3/2}$ state is approximately three times smaller than that of the $6P_{3/2}$ state~\cite{18,26}. Second, the Doppler width of the $6S_{1/2}\!\rightarrow\!7P_{3/2}$ transition is significantly larger; at $180^{\circ}\mathrm{C}$ it reaches $\Gamma_D/2\pi\simeq0.87\,\mathrm{GHz}$, about 2.5 times larger than for the $6S_{1/2}\!\rightarrow\!6P_{3/2}$ transition~\cite{26}. Third, the oscillator strengths of the $3\rightarrow2',3',4'$ transitions are nearly two orders of magnitude smaller than those for excitation to $6P_{3/2}$~\cite{26}. Consequently, the NC must be heated to temperatures of $\sim200^{\circ}\mathrm{C}$, as well as power of 456\,nm laser should be increased to achieve sufficient signal amplitude. At small $L$, additional vdW broadening causes spectral overlap, reducing the accuracy of frequency-shift determination.

For these reasons, particular attention was paid to the $4\rightarrow3',4',5'$ transitions. Their frequency separations are larger, and the amplitude of the strongest transition $A(4\rightarrow5')$ exceeds those of $A(4\rightarrow4')$ and $A(4\rightarrow3')$ by factors of 2.1 and 6.3, respectively. As a result, the determination of its vdW shift is less affected by neighboring transitions. As shown in Ref.~\cite{19}, results for the $4\rightarrow5'$ transition can be extrapolated to transitions originating from $F_g=3$ with an accuracy of about $10\%$.

For small vapor column thicknesses, the absorption coefficient follows $\alpha=\sigma N L$, where $\sigma\sim10^{-11}\,\mathrm{cm}^2$ is the absorption cross section, $N$ is the atomic density, and $L$ is the cell thickness~\cite{26}. Reliable detection at small $L$ therefore requires increased atomic density. For $L<80\,\mathrm{nm}$ and sufficiently high density, when the condition $N/k^3\gtrsim1$ is satisfied (with $k = 2\pi/\lambda$ the wavenumber), an additional red frequency shift due to Cs--Cs interactions appears, known as the Lorentz--Lorenz shift \cite{13},
\begin{equation}
\Delta\nu_{\mathrm{LL}}=-\frac{\pi N\gamma_N}{k^3},
\label{eq:LL}
\end{equation}
where $\gamma_N$ is the natural linewidth of the transition. For the $D_2$ line, $\gamma_N/2\pi=5.2\,\mathrm{MHz}$, while for the $6S_{1/2}\!\rightarrow\!7P_{3/2}$ transition $\gamma_N/2\pi=1.4\,\mathrm{MHz}$~\cite{20,27}. The total observed red shift is therefore
\begin{equation}
\Delta_T=\Delta_{\mathrm{vdW}}+\Delta\nu_{\mathrm{LL}},
\end{equation}
and $\Delta\nu_{\mathrm{LL}}$ must be subtracted to correctly determine the vdW coefficient $C_3$. Equation~(2) shows that the Lorentz--Lorenz frequency shift $\Delta\nu_{\mathrm{LL}}$ is substantially larger for the $6S_{1/2} \rightarrow 6P_{3/2}$ transition at $\lambda=852\,$nm, owing to its larger wavelength and natural linewidth\,\cite{13}.

Figure~4 illustrates the effect of vapor temperature on the Lorentz--Lorenz shift for the $4\rightarrow3',4',5'$ transitions at $\lambda=456\,\mathrm{nm}$ for $L=100\pm5\,\mathrm{nm}$ and laser power $3\,\mathrm{mW}$. Curves~1--5 correspond to temperatures of $200$, $245$, $270$, $295$, and $310^{\circ}\mathrm{C}$. The experimentally observed vdW shift is approximately $\Delta_{\mathrm{vdW}}\sim140\,$MHz, while the Lorentz--Lorenz contribution is $\Delta\nu_{\mathrm{LL}}\sim15\,\mathrm{MHz/Torr}$, reaching $\sim30\,\mathrm{MHz}$ at $310^{\circ}\mathrm{C}$. The reference SA spectrum is shown in the lower trace. Thus, for $T>200^{\circ}\mathrm{C}$, the Lorentz--Lorenz shift must be taken into account when extracting $C_3$.

Figure~5 presents transmission spectra of the $6S_{1/2}\!\rightarrow\!7P_{3/2}$ transition for different NC thicknesses. In Fig.~5(a), curves~1--3 correspond to $L=140\pm5$, $190\pm5$, and $230\pm5\,\mathrm{nm}$ at $190^{\circ}\mathrm{C}$. In Fig.~5(b), curves~1--3 correspond to $L=55\pm5$, $65\pm5$, and $100\pm5\,\mathrm{nm}$ at $230^{\circ}\mathrm{C}$. Owing to the relatively large $C_3$ coefficient, clear vdW broadening and a red shift of $\sim35\,\mathrm{MHz}$ are already observed at $L=230\,\mathrm{nm}$. At the same thickness, no measurable shift is observed for the $6S_{1/2}\!\rightarrow\!6P_{3/2}$ transition (see Fig.\,3). For $L>500\,\mathrm{nm}$, the vdW shift of the $6S_{1/2}\!\rightarrow\!7P_{3/2}$ transition becomes negligible.

Figure~6 shows the dependence of the vdW-induced red shift of the $4\rightarrow5'$ transition on $L$. At small $L$, spectral overlap between the broadened $4\rightarrow3'$, $4\rightarrow4'$, and $4\rightarrow5'$ transitions causes an additional red shift of the peak position, illustrated in the inset of Fig.~6.  This effect, referred to here as the overlapping red (OLR) shift, amounts to  approximately $30\,\mathrm{MHz}$ at $L = 90 \pm 5\,\mathrm{nm}$ and is subtracted from the measured shifts prior to the determination of $C_3$ using Eq.~(1), where the transmission spectrum is fitted by three Gaussian profiles corresponding to the $4 \rightarrow 3',4',5'$ transitions.

Figure~7 shows the resulting experimental values of the vdW coefficient $C_3$ as a function of $L$. The uncertainty arises from spectral overlap, Lorentz--Lorenz corrections, and the determination of the atom--surface distance entering Eq.~(1). A rapid decrease of the effective $C_3$ with decreasing $L$ is observed.

A similar reduction of the vdW coefficient with decreasing atom--surface distance was previously reported for the $5S_{1/2} \rightarrow 5P_{1/2}$ transition in Rb and the $6S_{1/2} \rightarrow 6P_{1/2}$ transition in Cs~\cite{13}. This behavior was theoretically predicted in Refs.~\cite{28,29} and attributed to retardation effects. In this interpretation, the measured signal reflects the difference between the vdW potentials of the excited and ground states. While the excited-state potential remains well approximated by $-C_3/z^3$, the ground-state potential transitions to a faster-decaying $-C_4/z^4$ form beyond distances of order $50\,\mathrm{nm}$, effectively enhancing the measured difference. Although this explanation was developed for the $6S\!\rightarrow\!6P$ transition, similar considerations may apply to the $6S\!\rightarrow\!7P$ transition.

Another possible contribution to the apparent reduction of the vdW effect at small $L$ arises from photoelectric charging of the NC windows. It has been shown that irradiation of Cs vapor cells with $\lambda\sim453\,$nm  light can induce photoemission of electrons from the walls, creating electric fields that produce blue frequency shifts~\cite{30,31}. In the present experiment, the $456\,\mathrm{nm}$ radiation itself may induce a similar effect. In Ref.~\cite{32}, application of a static electric field of $15\,\mathrm{kV/cm}$ to a centimeter-long Rb cell resulted in a red shift of $\sim20\,\mathrm{MHz}$ for the $5S_{1/2}\!\rightarrow\!5P_{3/2}$ transition, in agreement with theory.

Because of stronger spectral overlap and comparable transition amplitudes, the vdW effect was quantitatively analyzed primarily for the $4\rightarrow3',4',5'$ transitions. Nevertheless, Fig.~8 demonstrates that the $3\rightarrow2',3',4'$ transitions also exhibit vdW-induced red shifts for $L=60\pm5$, $70\pm5$, $90\pm5$, and $100\pm5\,\mathrm{nm}$ at $230^{\circ}\mathrm{C}$ and laser power $3\,\mathrm{mW}$. Importantly, this red shift contrasts with earlier SR measurements for the $6S_{1/2}\!\rightarrow\!6P_{3/2}$ transition, where an unexpected blue shift was reported for analogous transitions~\cite{7}. The spectra in Fig.~8 show the total red shift, including contributions from $\Delta_{\mathrm{vdW}}$, $\Delta\nu_{\mathrm{LL}}$, and the OLR shift.

Finally, we note that there is currently no simple and widely accessible experimental method described in the literature for studying vdW effects in the $6S_{1/2}\!\rightarrow\!7P_{3/2}$ transition that allows direct comparison with nanocell results. Recently developed glass nanocells~\cite{33,34} are unsuitable for this transition because at the required temperatures ($>190^{\circ}\mathrm{C}$) chemical interaction of hot Cs vapor with glass leads to rapid cell degradation. In contrast, the sapphire-based nanocells used here can operate at temperatures up to $450^{\circ}\mathrm{C}$~\cite{12,13,14,15,16,17}. Saturated absorption spectroscopy in millimeter-scale cells has recently been applied to the $6S_{1/2}\!\rightarrow\!7P_{1/2}$ transition, achieving short-term frequency stability below $2\times10^{-13}$~\cite{35,36}. For submicron cell thicknesses, however, vdW interactions must be taken into account. We anticipate that the present results will stimulate further theoretical development of vdW and Casimir--Polder interactions for the Cs $6S_{1/2}\!\rightarrow\!7P_{3/2}$ transition.

\section{Conclusion}

A homemade optical cell fabricated from technical sapphire, featuring a wedge-shaped gap between the inner window surfaces smoothly varying from $50$ to $2000\,\mathrm{nm}$, was employed. Such a cell, hereafter referred to as a nanocell (NC), was filled with cesium atomic vapor. Using this NC, the influence of a closely spaced dielectric surface on the transmission spectrum of Cs atoms at the $6S_{1/2}\!\rightarrow\!7P_{3/2}$ ($D_2$) transition was experimentally investigated for the first time in the thickness range $50$--$250\,\mathrm{nm}$.

For atom--surface distances $L\lesssim 300\,$nm, a large red frequency shift of the atomic transitions was observed, arising from the van der Waals (Casimir--Polder) interaction between Cs atoms and the sapphire surface of the nanocell. An additional red shift was found at elevated cesium vapor pressures, with a measured value of approximately $15\,\mathrm{MHz/Torr}$. This contribution must be taken into account for the correct determination of the van der Waals interaction coefficient $C_3$.

By recording transmission spectra of the nanocell at different thicknesses, the vdW coefficient for the Cs $6S_{1/2}\!\rightarrow\!7P_{3/2}$ transition was determined experimentally for the first time. The measured values lie in the range
\begin{equation*}
C_3 = (2\text{--}20)\,\mathrm{kHz}\,\mu\mathrm{m}^3 \end{equation*}
A comparative study of vdW-induced frequency shifts for the $6S_{1/2}\!\rightarrow\!6P_{3/2}$ and $6S_{1/2}\!\rightarrow\!7P_{3/2}$ transitions was also carried out. In particular, it was shown that at a nanocell thickness of approximately $200\,\mathrm{nm}$, a pronounced red shift is clearly observed for the $6S_{1/2}\!\rightarrow\!7P_{3/2}$ transition, whereas no measurable frequency shift is detected for the $6S_{1/2}\!\rightarrow\!6P_{3/2}$ transition. This behavior is attributed to the substantially smaller vdW coefficient $C_3$ associated with the latter transition.

The results obtained are of importance for the development of miniature sensors based on atomic vapors and, in particular, for the realization of compact frequency references exploiting atomic transitions in the blue spectral region~\cite{14}.

\section*{Author Contributions}
Armen Sargsyan: Investigation, Writing -- review \& editing, Funding acquisition. Anahit Gogyan: review \& editing, Visualization, Methodology, Formal analysis. David Sarkisyan: review \& editing, Writing -- original draft, Visualization, Validation, Supervision
\section*{Acknowledgments}
The authors are grateful to Tigran Vartanyan and Athanasios Laliotis for helpful discussions. 

\section*{Funding}
The research was supported by the Higher Education and Science Committee of MESCS RA (Research project N25RG-1C008).

\end{document}